\documentclass[conference]{IEEEtran}
\IEEEoverridecommandlockouts
\usepackage{cite}
\usepackage{amsmath,amssymb,amsfonts}
\usepackage{algorithmic}
\usepackage{graphicx}
\usepackage{textcomp}
\usepackage{xcolor}
\usepackage{float}
\usepackage{makecell}
\usepackage{subfigure}
\usepackage{hyperref}
\graphicspath{{images/}}
\def\BibTeX{{\rm B\kern-.05em{\sc i\kern-.025em b}\kern-.08em
    T\kern-.1667em\lower.7ex\hbox{E}\kern-.125emX}}

\begin{document}

\title{Geodesic trajectories for binary systems of supermassive black holes (SMB)\\
}

\author{\IEEEauthorblockN{Fabian A. Portilla}
\IEEEauthorblockA{\textit{Departamento de Fisica} \textit{Universidad del Cauca}\\
Popay\'an, Colombia \\
fabianportilla@unicauca.edu.co}
}

\maketitle

\begin{abstract}
  During this work, it is considered a binary system of 
  supermassive rotating black holes; first, it is employed the
  concept of weak field limit to develop a metric tensor 
  $g_{\mu\nu}$ that describes the geometry of the spacetime, 
  it introduced an approximation in which the second black 
  hole is coupled to the system through a perturbation tensor $f^{*}_+$,
  consequently, Ricci scalar curvature is determined at a distance 
  of 120 AU and whose value is $-1.97\cdot 10^{-34}$ $m^{-2}$,
  it is employed a black hole type Sagittarius A* to make the numerical 
  calculations; the negative Ricci scalar curvature states that 
  the tensor $f^{*}_+$ does not change the topological 
  properties of Kerr's solution. \\
  
  From the metric tensor developed and the scalar of curvature the geodesic 
  trajectories are derived; they determine an orbit with a perigee of 116.4AU 
  and an apogee of 969.67AU, the orbit has a precession of 
  77.8 seconds per year; and the precession is determined by 
  the rotation of the black holes besides the angular momentum 
  that is the classical parametrization; finally, the average 
  energy is defined by the equation 
  $\langle E \rangle_T = c^2 \sqrt{(GMa_{major} (1-\epsilon^2 )/2a^2 )}$, 
  this expression parametrizes the energy per orbit in function of the rotation 
  of the black holes,
  and whose value is $-2.99 \cdot 10^13$  J/kg, this value is one 
  order of magnitude higher than Newtonian energy.\\[0.2cm]
\end{abstract}


\section{Introduction}

The study of gravity began in 1687 when Sir Isaac Newton 
published his Principia \cite{Principia}, where law of 
universal gravitation was stated, in this conception, 
gravity is a force acting at distance \cite{Sean,Cosmology}, 
it predicted a scalar potential field that describes the motion 
of any test particle in the universe; after, in 1916, Albert 
Einstein published his general theory of relativity; where the 
gravity is stated as a consequence of the geometry of spacetime,
 Einstein set the Newton's theory as a linear approximation of 
 the Einstein field equation; and the geometry of spacetime is 
 described as a manifold $R^{\{3,1\}}$ ; in 1964, Roy Kerr found 
 a solution that predicted the existence of rotating black holes,
  these solutions correctly describe the spacetime geometry, 
  however, their mathematical structure impedes to obtain 
  analytical solutions.\\

In 2022, the Event Horizon telescope detected the first image 
of a real black hole called \textbf{Sagittarius A*} \cite{detec}; when the first black hole 
was observed at the center of the milky way, the mass and size 
of Sagittarius A* were estimated employing the motion of the 
stars in the Sagittarius constellation, it means that the 
theoretical models that describe the geodesics of spacetime
 are written in terms of the mass and spin of the black 
 hole; the computational simulations are very powerful tools 
 to analyze the orbits of the stars, in this sense, numerical 
 methods such as finite differences or the entire branch of 
 numerical relativity allow to solve the Einstein field equation 
 at each point of spacetime, however, it limits to extrapolate 
 the information beyond the data obtained from the observations.\\

The spacetime is understood as a dynamical structure whose 
topology is the same than $R^{\{3,1\}}$  \cite{Sean,Cosmology}; 
the mass and the energy deform the spacetime tissue  and 
produce folds in the universe itself, hence, the motion of 
the planets and stars is product of these folds \cite{Weinberg}, 
the spherical symmetry of the stars and planets make that the 
deformations of the spacetime have spherical symmetry. When a 
stellar system is conformed for more than one object, the 
central body often contains between the 80-95$\%$ of the mass 
of the system, as a consequence, the dynamics of the rest of 
the objects in the system can be deduced analyzing the central 
mass; due to the astronomical observations, it is known that 
around of 80$\%$ of the systems are binary \cite{detec,Thorme}, 
and the Einstein equations can be worked to analyze both objects 
at the same time, nonetheless, these solutions are not exact 
instead they are approximations applied to systems of low 
energies, thus the solutions can not predict many features of 
the system such as evolution of the precession, and conserved 
quantities.\\

The Einstein equation is not a linear relation, it means that 
the general metric of a system is not the superposition of the 
solutions for each body that composes the system; to construct 
the metric tensor the Einstein field equation must be solved 
considering all bodies at the same time, if the spacetime is 
thought as a dynamical structure each system will have a 
different geometry depending on the distribution of mass and 
energy, nonetheless, the general theory of relativity is a 
topological theory as well \cite{geometry}, hence, the 
spacetime $R^{\{3,1\}}$  is isomorphic to $R^3$, it is a special
 case where the Newtonian mechanics works, when the systems 
 are worked in the weak field limit the dynamics of 
 any body is determined by the Newtonian potential $\Phi$, 
 hence, when weak field limit is implemented, all systems can be worked with the same potential $\Phi$.

\section{Discussion}

We consider a system composed by two supermassive black holes 
(SMB), and whose physical characteristics are the same than 
Sagittarius A* located at the center of Milky Way \cite{detec,detection},
we commence to determine a linear metric tensor that considers the 
presence of both black holes.

\subsection*{Metric tensor in weak field limit approximation} \label{MT}

A system of two rotating supermassive black holes does not have exact solutions from the 
Einstein's field equation, hence, we approximate the metric of 
the system for a region of the spacetime where the gravitational field is 
linear \cite{Sean}; in large distances of $r$, Kerr's metric  approximates as follows \cite{Cosmology}:

\begin{equation}
  \begin{split}
    g_{kerr} \cong &-(1-2\Phi)dt^2 + (1-2\Phi)^{-1}dr^2 + r^2d\theta^2\\
    & + (r^2sin^2\theta -(1-2\Phi)a^2sin^4\theta)d\phi^2 \\
    &-2\Phi a sin^2\theta(dtd\phi + d\phi dt)
  \end{split}
  \label{eqn 2.1}
\end{equation}

The equation (\ref{eqn 2.1}) is written in terms of $\Phi$ that 
in a
linear approximation corresponds to the \textit{Newton 
potential}, it is valid in the range of values where $r>>a/c$ \cite{Sean,Cosmology}, 
consequently, as $\Phi$ is  linear,
the resulting field is the linear superposition of the field 
produced by 
each black hole \footnote{We work in units with $c=1$, 
the rest of the numerical values are given in I.S. units.}; 
we find the form of the potential $\Phi$ considering one 
of the 
 black holes located at the origin of the coordinate system 
 $r=0$,
 and the second black hole at $x = -l$, we apply a 
 Taylor expansion 
 and ignore higher order terms, the following form is obtained.\\
 
 \begin{eqnarray}
  \Phi(r,\theta,\phi) = \frac{2GM}{r} - \frac{psin\theta cos\phi}{r^2}
  \label{eqn 2.2}
 \end{eqnarray}

 The equation (\ref{eqn 2.2}) introduces the term $p = GMl$ that will be 
 called \textit{mass dipole moment}; it is only a quantity that 
 determines the distribution of mass in spacetime; employing 
 the equation (\ref{eqn 2.2}) we propose the metric tensor of the 
 system as:

\begin{equation}
  g_{\mu\nu} = g_{kerr}(\phi_0) + f^{*}_+(\phi_p)
  \label{eqn 2.4}
 \end{equation}

 \begin{equation}
    \begin{split}
      g_{kerr}=& -\left(1-\frac{4GM}{r}\right)dt^2 + \left(1-\frac{4GM}{r}\right)^{-1}dr^2\\
      & + (r^2sin^2\theta -\left(1-\frac{4GM}{r}\right)a^2sin^4\theta)d\phi^2 \\
      &+ r^2d\theta^2  -\left(\frac{4GM}{r}\right) a sin^2\theta(dtd\phi + d\phi dt)
    \end{split}
    \label{eqn 2.5}
\end{equation}

\begin{equation}
  \begin{split}
    f^{*}_+=& -\left(\frac{2psin\theta cos\phi}{r^2}\right)\left[dt^2 + dr^2 + a^2sin^4\theta d\phi^2 \right]\\
    &--\left(\frac{2psin\theta cos\phi}{r^2}\right)\left[asin^2(\theta)(dtd\phi + d\phi dt)\right]
  \end{split}
  \label{eqn 2.6}
\end{equation}

  The term $g_{kerr}(\phi_0)$ takes the same form shown in the equation 
(\ref{eqn 2.1}) except that $\phi_0$ is the potential produced by the two 
black holes as whether they were located at a single point, in contrast, the tensor 
shown in the equation (\ref{eqn 2.6}) states an small perturbation 
in the background metric \cite{Sean}; this tensor 
depends on $1/r^2$, and as a consequence, $\vert f^{*}_+\vert << \vert g_{kerr}\vert$; the 
tensor (\ref{eqn 2.5}) is an exact solution of Einstein field equation \cite{Sean}, while the tensor (\ref{eqn 2.6}) is not, hence, in general the 
equation (\ref{eqn 2.4}) does not satisfy the Einstein field equation, nonetheless, 
we will keep this term because it considers the separation between the black holes.\\

The approximation made in the equation (\ref{eqn 2.4}) impedes to obtain
geodesic curves valid in all regions of spacetime, clearly at the origin 
the tensor $f_{\mu\nu}$ diverges to $\pm \infty$, hence, in 
general, the geodesics will be inconsistent, but at larger distances 
from the origin, the geometric perturbation is small enough such that 
the geodesic equations are determined by the equation (\ref{eqn 2.5})
with linear corrections given by the equation (\ref{eqn 2.6}); in the astronomical 
observations, the binary systems of black holes have not been found, however, 
we will use the orbit of the star S2 in the Sagittarius constellation to test the model 
developed in this paper \cite{detec}, although the system is not binary, 
the equations derived from the model must determine the trajectory 
followed by S2.

\subsection*{Ricci tensor convergence } \label{RT}

The metric tensor considering a vacuum case must 
satisfy the condition that \cite{Cosmology}.

\begin{equation}
  R_{\mu\nu}(g_{\mu\nu}) = 0
\end{equation}

The tensor (\ref{eqn 2.4}) satisfies this condition because 
it is an exact solution \cite{Cosmology}, however, we are 
interested in the variations of the Ricci tensor produced by 
$f^{*}_{+}$, as the solution is only an approximation, 
a real system is introduced to obtain numerical values, consider 
a binary system of black holes similar to Sagittarius A* \cite{detec, detection},
but we restrict the system to be stationary and avoid the gravitational radiation
by the motion of spacetime \footnote{The gravitational radiation implies an approximation that can be consulted in chapter 7 \cite{Sean}}\cite{Sean, Cosmology}; the terms of $R_{\mu\nu}(f^{*}_+)$
decays in the form $1/r^n$, hence we will compute the numerical 
value at some fixed distance, the numerical value for higher distances 
will be smaller and converge to zero; for this propose, we consider 
the orbit of the star S2 at Sagittarius constellation. \footnote{The star S2 has a perigee of 
120AU, and an apogee of 970AU}

\begin{equation}
  \langle R\rangle =
  \begin{pmatrix}
           -2.91 \cdot 10^{-18} & 0 & 0 & -2.41 \cdot 10^{-14} \\
           0 & -4.81 \cdot 10^{-31} & 0 & 0 \\
           0 & 0 & 5.37 \cdot 10^{-6} & 0 \\
           -2.41 \cdot 10^{-14} & 0 & 0 & 5.39 \cdot 10^{-6}
  \end{pmatrix}
  \label{eqn 2.8}
\end{equation}

  The above square representation optimizes the 
values of $\theta, \phi$, 
such that the equation (\ref{eqn 2.8}) has the 
maximum variation with respect to 
zero, for values $r >> 120AU$, the tensor will 
have smaller values than those shown in the
equation (\ref{eqn 2.8}) \cite{Sean}, the highest 
values in the Ricci 
tensor are given in the position $d\theta$ 
and $d\phi$, the equation (\ref{eqn 2.6}) shows
 that the perturbation tensor depends on $a$, 
this parameter has 
a value of $(2.23 \pm 0.41)\cdot 10^{9}m$, 
consequently, it is responsible 
for the variations of the components $\langle R_{33} \rangle$ and 
$\langle R_{44} \rangle$.

\subsection*{Geodesic trajectories} \label{GT}

We begin to derive the equations of motion of any test particle 
moving in the spacetime described by the tensor (\ref{eqn 2.4}), 
particularly, we can state that the background metric is the 
Kerr's metric, hence solutions agree with Kepler's laws must be 
predicted by the equation (\ref{eqn 2.4}) \cite{Cosmology, Sean};
we begin by looking at 
the conserved quantities; the metric tensor 
(\ref{eqn 2.4}) is time-independent, which means that there exists a
\textit{Killing vector} describing the conservation of energy 
\cite{geometry, diff}, nevertheless, the expression for the 
conservation of the energy is an approximation that will be 
stated as follows \footnote{All conserved quantities by the 
metric tensor are defined by an exact solution of Einstein Field 
equation \cite{Sean}.}

\begin{equation}
  \begin{split}   
  E = \left(1 - \frac{4GM}{r}\right)\frac{dt}{d\lambda} + \\
  \frac{2psin(\theta)cos(\phi)}{r^2}\frac{dt}{d\lambda} + E^{(2)} ..
\end{split}
  \label{eqn 2.9}
\end{equation}

The second term of the equation (\ref{eqn 2.9}( and the higher order terms 
$E^{(2)}...$ are introduced by the metric tensor $f^{*}_{+}$, however, 
the higher order terms vanish when $r$ increases; using the Killing vector 
for \textit{t-coordinate}, we can obtain the variations of the equation (\ref{eqn 2.9})
respect to zero, the Killing vector related to the total energy of the system 
is given by: \cite{Sean,geometry}.

\begin{equation}
  K_\mu = \left(1-\frac{4GM}{r} + \frac{2psin(\theta)cos(\phi)}{r^2},0,0,0\right)
\label{eqn 2.10}
\end{equation}

The killing vector is given by the coefficient of $g_{tt}$, because 
the metric tensor is independent of time \cite{Sean}, for an exact solution 
the Killing equation is equal to zero, nonetheless, in our approximation it is not the case,
the numerical values of the killing equations at $120AU$ are shown in 
the following matrix:

\begin{equation}
  \nabla_\nu K_\mu =
\begin{pmatrix}
  0 & 2.51 \cdot 10^{-10} & 0 & 0\\
  2.51 \cdot 10^{-10} & 0 & 0 & 1.833\\
  0 & 0 & 0 & 0\\
  0 & 1.833 & 0 & 0
\end{pmatrix}
\label{eqn 2.11}
\end{equation}
 
The component $\nabla_rK_t$ varies by the presence of $\phi_p$, and the
other terms different from zero are due to the presence of angular variables, 
however the order of the variation $10^{-10}$ allows to approximate 
the energy as a constant of motion given by the equation (\ref{eqn 2.9}).

\begin{equation}
  \begin{split}
    L = & (r^2-a^2\left( 1-\frac{4GM}{r} + \frac{2psin(\theta)cos(\phi)}{r^2}\right))\frac{d\phi}{d\lambda} \\
    &= (r^2-a^2\mathfrak{F})\frac{d\phi}{d\lambda}
  \end{split}
  \label{eqn 2.12}
\end{equation}

  The equation (\ref{eqn 2.12}) is a linearization of the angular momentum, 
in this system by the separation of the black holes, the 
the system does not preserve the angular momentum as a constant of motion,
thus, the equation (\ref{eqn 2.12}) is an approximation made to reduce the 
complexity of the differential equations \cite{diff}, we can obtain the differential equations that parametrize any curve in the spacetime as follows.

\begin{equation}
  \begin{split}
  E^2 =& \left(\frac{dr}{d\lambda}\right)^2 + \left[\epsilon -\epsilon\frac{4GM}{r} + \frac{L^2}{r^2} - \frac{4GML^2}{r^3}\right] \\
  & + \left[\frac{2p\epsilon cos\phi}{r^2} - \frac{4GM}{r^3}(2ELa) + \frac{2pcos\phi}{r^4}(L^2 + 2ELa)\right] 
  \end{split}
  \label{eqn 2.13}
\end{equation}

\begin{equation}
  \left(\frac{dr}{d\lambda}\right)^2 + V_{Sch}(r) + V_{pert}(r) = E^2
  \label{eqn 2.14}
\end{equation}

The equation (\ref{eqn 2.13}) is the general parametrization of a curve 
in the spacetime given by the metric tensor $g_{\mu\nu}$, it is not 
an exact solution, in the sense that $r^2 + a^2 \approx r^2$, under this 
approximation, we see that the equation \ref{eqn 2.14} tales the form of the
Schwarzschild differential equation for geodesic curves, nonetheless, $V_{pert}(r)$ 
is not only the contribution of the perturbation $f_{\mu\nu}$ but the rotation of 
the black holes, that is a geometrical characteristic that is not present in 
the Schwarzschild spacetime, in this sense, we have got to incorporate the second 
black hole to the system and the rotation of the black hole while we keep 
an exact parametrization of the trajectories \cite{Sean,Weinberg}; we consider that our solution is given by the summation 
of three orders of approximation, $r^{(0)},r^{(1)},r^{(2)}$ 
hence, the trajectory is given by:

\begin{equation}
  r(\phi) = \frac{L^2}{O(r)}
  \label{eqn 2.15}
\end{equation}
\begin{equation}
  \begin{split}
    O(r) = &2GM(1 + \epsilon cos(\beta - 1) + \left(1+\frac{\epsilon^2}{3}\right)\beta^2\epsilon\phi sin(\phi) -\\
    & \frac{\beta^2\epsilon\phi^2}{2}cos(\phi) )
  \end{split}
\end{equation}

In the equation (\ref{eqn 2.15}), the first two terms of the denominator
are the responsible of determining an elliptical trajectory and the precession
of the orbit, in the equation \ref{eqn 2.15}, $\epsilon$ determines 
the eccentricity of the orbit, for numerical calculations we will use the value of $0.88$; on the other hand, the $\beta$ parameter
is given by:

\begin{equation}
  \beta = \frac{12G^2M^2}{L^2} + \frac{24G^2M^2Ea}{L^3}
\end{equation}

The precession of the orbit is given by $\beta$ parameter, for our system we have a 
value of $78.7$ \textit{seconds per year}; the second order of 
approximation $r^{(2)}$ allows to determine the
average energy stored in an orbit.

\begin{equation}
  \langle E \rangle_T = -c^2\sqrt{\frac{GMa_{major}(1-\epsilon^2)}{2a^2}}
  \label{eqn 2.17}
\end{equation}

The equation (\ref{eqn 2.17}) is an expression that considers the 
rotation of the black hole as a contribution to the energy of 
the test particle, the term $a_{major}$ differentiates the orbits,
considering a perigee of $120AU$ the energy has a value of 
$\langle E \rangle_{T0} = -2.99 \cdot 10^{13}J/kg$, we plot the 
equation \ref{eqn 2.15} to represent the trajectory.\\

\begin{figure}[ht]
  \begin{center}
       \includegraphics[width=8.5cm, height=7cm]{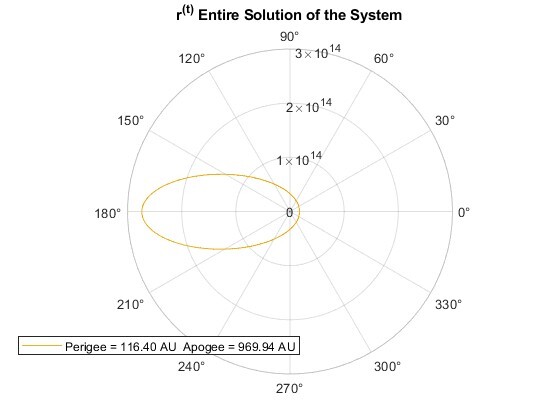}
       \caption{Elliptical trajectory described by a star with eccentricity $\epsilon = 0.88$ and semi major axis $120AU$}
       \label{figure 1}     
      \end{center}  
\end{figure}

\begin{figure}[ht]
  \begin{center}
      \subfigure[Projection of the geodesic trajectories]{
       \includegraphics[width=7cm, height=5.5cm]{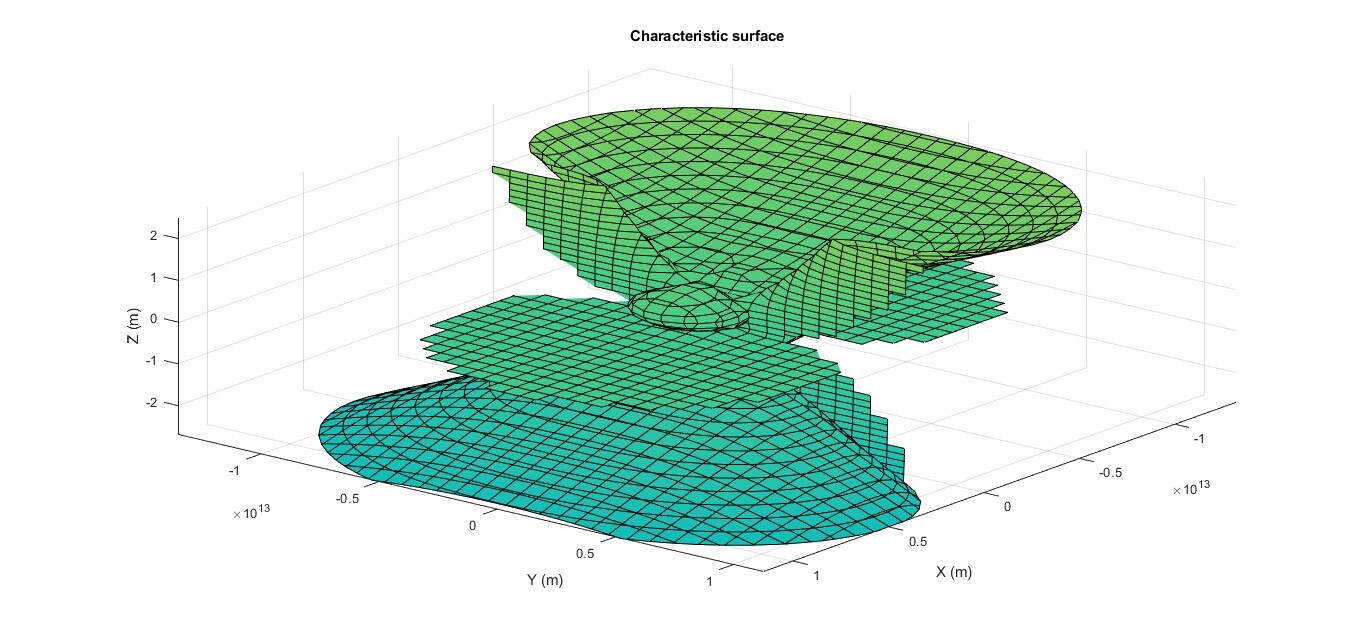}
       \label{figure 2}}
       \subfigure[embedded surfaces of the spacetime]{
       \includegraphics[width=7cm, height=5.5cm]{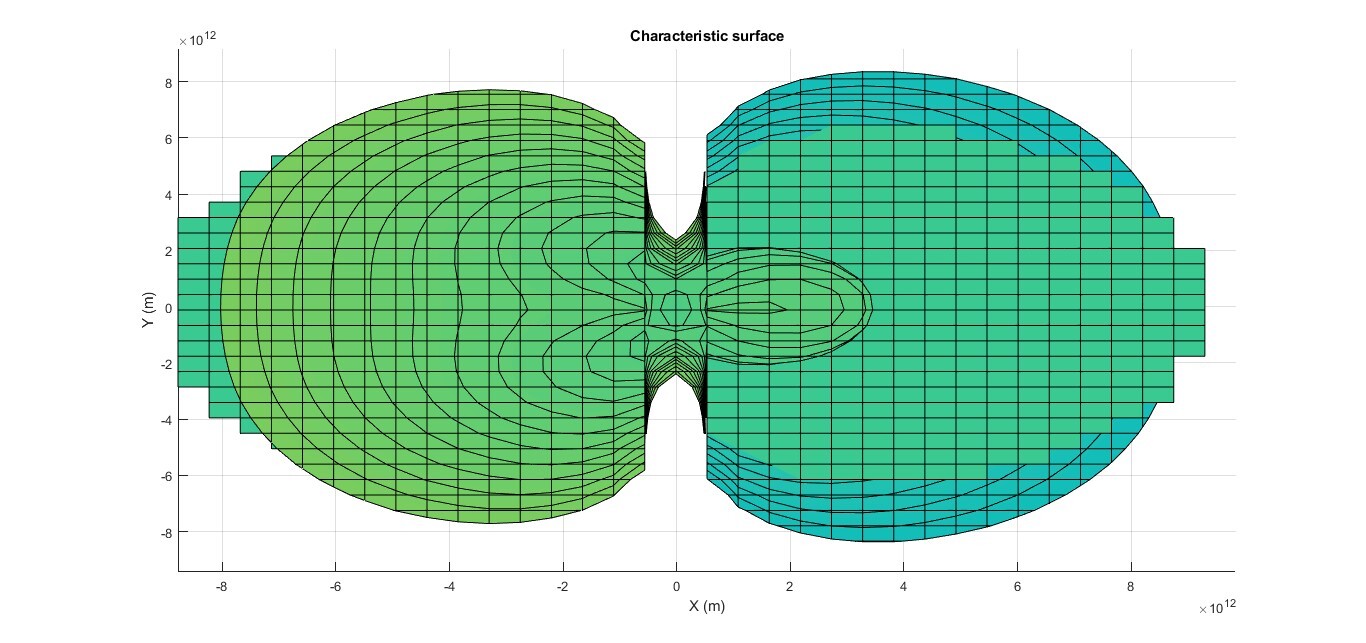}
       \label{figure 3}}
       \caption{Characteristic surfaces created from  the metric tensor $g_{\mu\nu}$, surfaces that represent the geodesic trajectories projected in a 3-dimensional plane.}
      \end{center}  
\end{figure}
  
  The figure \ref{figure 1} represents the position in function of the 
angle $\phi$, in fact, it is a closed trajectory following the 
Kepler's laws \cite{Cosmology}, the rest of terms in the denominator modify
the precession of the orbit, we can plot the projection of the 
surfaces given by the metric tensor (\ref{eqn 2.5}), the figure \ref{figure 2}
are $t-constant $ surfaces projected in the plane perpendicular to 
the angular momentum axis, in the weak field limit, the orbits tend to be ellipses, and 
near to the origin we distinguish two regions that refer to the position 
of the black holes, on the other hand, the figure \ref{figure 3} represents 
the embedded surface in the spacetime, it is curved towards the origin and represents
the attraction to the gravitational sources \cite{Cosmology}, however, the 
\textit{z-constant slices} are not circular but elliptical, near to the origin 
we have an approximated representation of the system, the no connected 
regions mean that the tensor (\ref{eqn 2.4}) fails to describe the spacetime near 
to the black holes.

\section*{Conclusions}

In the case of a coupled system, we established  
the distribution of the masses and the conditions of symmetry as follows; 
the black holes are strictly static with each other to avoid 
the temporal evolution of spacetime; hence, in weak field limit 
the metric tensor is stationary; on the other hand, the 
non-diagonal terms of Kerr's metric tensor are deduced 
theoretically only in the case where the orbit is defined in 
a plane containing the accretion disk; the approximation determines that the 
relativistic scalar potential $\mathfrak{F} = 1 - 2\Phi$ that determines
the effect of the masses in the curvature of the spacetime
has a value of $1.000011$; at a distance of $120AU$ the Ricci tensor has numerical values that oscillate 
between $10^{-6}m^{-2}$ and $10^{-31}m^{-2}$; on the other hand, 
the scalar curvature $R$ at 120AU has a value of 
$-1.97 \cdot 10^{-34}m^{-2}$, the negative value of the curvature
shows that the metric tensor $g_{\mu\nu}$ is topologically admissible to 
the manifold $R^{\{3,1\}}$.\\

The trajectories obtained in the weak field limit are ellipses from a 
distance of $120AU$ up to $+\infty$, the solution fixes 
the nearest point at $116.4AU$, and the apsis is at $969.97AU$; 
the precession of the orbit has a value of 
$78.7$ \textit{seconds per year},
in this system the precession is a consequence of the deformation 
produced by the masses, and $1.3\%$ of the value of such a 
precession is given by the rotation of the black holes;
the differential equations of the geodesic curves allow to 
parametrize an equation for the average energy per unit orbit
that is given by $\langle E\rangle_T=-c^2\sqrt{GMa_{maj}(1-\epsilon^2)/2a^2}$,
 the energy 
per orbit has a value of 
$-2.99 \cdot 10^{13}J/kg$ which presents one order magnitude 
higher than the \textit{Newton's energy}.\\

\end{document}